\input harvmac
\input epsf

\font\cmss=cmss10
\font\cmsss=cmss10 at 7pt

\def\IZ{\relax\ifmmode\mathchoice
{\hbox{\cmss Z\kern-.4em Z}}{\hbox{\cmss Z\kern-.4em Z}}
{\lower.9pt\hbox{\cmsss Z\kern-.4em Z}}
{\lower1.2pt\hbox{\cmsss Z\kern-.4em Z}}\else{\cmss Z\kern-.4em
Z}\fi}

\newcount\figno
\figno=0 
\def\fig#1#2#3{
\par\begingroup\parindent=0pt\leftskip=1cm\rightskip=1cm\parindent=0pt
\baselineskip=11pt
\global\advance\figno by 1
\midinsert
\epsfxsize=#3
\centerline{\epsfbox{#2}}
\vskip 12pt
{\bf Fig.\ \the\figno: } #1\par
\endinsert\endgroup\par
}

\nref\KlebanovQA{
I.~R.~Klebanov,
``String theory in two-dimensions,''
arXiv:hep-th/9108019.
%%CITATION = HEP-TH 9108019;%%
}

\nref\GinspargIS{
P.~Ginsparg and G.~W.~Moore,
``Lectures On 2-D Gravity And 2-D String Theory,''
arXiv:hep-th/9304011.
%%CITATION = HEP-TH 9304011;%%
}

\nref\PolchinskiMB{
J.~Polchinski,
``What is string theory?,''
arXiv:hep-th/9411028.
%%CITATION = HEP-TH 9411028;%%
}

\nref\ShenkerUF{
S.~H.~Shenker,
``The Strength Of Nonperturbative Effects In String Theory,''
RU-90-47,
presented at the Cargese Workshop on Random Surfaces, Quantum Gravity
and Strings, Cargese, France, May 28 - June 1, 1990.
}

\nref\PolchinskiFQ{
J.~Polchinski,
``Combinatorics Of Boundaries In String Theory,''
Phys.\ Rev.\ D {\bf 50}, 6041 (1994)
[arXiv:hep-th/9407031].
%%CITATION = HEP-TH 9407031;%%
}

\nref\McGreevyKB{
J.~McGreevy and H.~Verlinde,
``Strings from tachyons: The $c = 1$ matrix reloaded,''
arXiv:hep-th/0304224.
%%CITATION = HEP-TH 0304224;%%
}

\lref\DasKA{
S.~R.~Das and A.~Jevicki,
``String Field Theory And Physical Interpretation Of D = 1 Strings,''
Mod.\ Phys.\ Lett.\ A {\bf 5}, 1639 (1990).
%%CITATION = MPLAE,A5,1639;%%
}

\lref\DiFrancescoUD{
P.~Di Francesco and D.~Kutasov,
``World sheet and space-time physics in two-dimensional (Super)string theory,''
Nucl.\ Phys.\ B {\bf 375}, 119 (1992)
[arXiv:hep-th/9109005].
%%CITATION = HEP-TH 9109005;%%
}

\lref\MooreZV{
G.~W.~Moore, M.~R.~Plesser and S.~Ramgoolam,
``Exact S matrix for 2-D string theory,''
Nucl.\ Phys.\ B {\bf 377}, 143 (1992)
[arXiv:hep-th/9111035].
%%CITATION = HEP-TH 9111035;%%
}

\lref\DijkgraafHK{
R.~Dijkgraaf, G.~W.~Moore and R.~Plesser,
``The Partition function of 2-D string theory,''
Nucl.\ Phys.\ B {\bf 394}, 356 (1993)
[arXiv:hep-th/9208031].
%%CITATION = HEP-TH 9208031;%%
}

\lref\JevickiZG{
A.~Jevicki and T.~Yoneya,
``A Deformed matrix model and the black hole background in
two-dimensional string theory,''
Nucl.\ Phys.\ B {\bf 411}, 64 (1994)
[arXiv:hep-th/9305109].
%%CITATION = HEP-TH 9305109;%%
}

\lref\DemeterfiCM{
K.~Demeterfi, I.~R.~Klebanov and J.~P.~Rodrigues,
``The Exact S matrix of the deformed c = 1 matrix model,''
Phys.\ Rev.\ Lett.\  {\bf 71}, 3409 (1993)
[arXiv:hep-th/9308036].
%%CITATION = HEP-TH 9308036;%%
}

\lref\CallanUB{
C.~G.~Callan, I.~R.~Klebanov, A.~W.~Ludwig and J.~M.~Maldacena,
``Exact solution of a boundary conformal field theory,''
Nucl.\ Phys.\ B {\bf 422}, 417 (1994)
[arXiv:hep-th/9402113].
%%CITATION = HEP-TH 9402113;%%
}

\lref\NatsuumeSP{
M.~Natsuume and J.~Polchinski,
``Gravitational Scattering In The C = 1 Matrix Model,''
Nucl.\ Phys.\ B {\bf 424}, 137 (1994)
[arXiv:hep-th/9402156].
%%CITATION = HEP-TH 9402156;%%
}

\lref\PolchinskiJP{
J.~Polchinski,
``On the nonperturbative consistency of d = 2 string theory,''
Phys.\ Rev.\ Lett.\  {\bf 74}, 638 (1995)
[arXiv:hep-th/9409168].
%%CITATION = HEP-TH 9409168;%%
}

\lref\DharGW{
A.~Dhar, G.~Mandal and S.~R.~Wadia,
``Discrete state moduli of string theory from the C=1 matrix model,''
Nucl.\ Phys.\ B {\bf 454}, 541 (1995)
[arXiv:hep-th/9507041].
%%CITATION = HEP-TH 9507041;%%
}

\nref\MartinecKA{
E.~J.~Martinec,
``The annular report on non-critical string theory,''
arXiv:hep-th/0305148.
%%CITATION = HEP-TH 0305148;%%
}

\nref\KlebanovKM{
I.~R.~Klebanov, J.~Maldacena and N.~Seiberg,
``D-brane decay in two-dimensional string theory,''
JHEP {\bf 0307}, 045 (2003)
[arXiv:hep-th/0305159].
%%CITATION = HEP-TH 0305159;%%
}

\nref\McGreevyEP{
J.~McGreevy, J.~Teschner and H.~Verlinde,
``Classical and quantum D-branes in 2D string theory,''
arXiv:hep-th/0305194.
%%CITATION = HEP-TH 0305194;%%
}

\nref\SchomerusVV{
V.~Schomerus,
``Rolling tachyons from Liouville theory,''
arXiv:hep-th/0306026.
%%CITATION = HEP-TH 0306026;%%
}

\nref\AlexandrovNN{
S.~Y.~Alexandrov, V.~A.~Kazakov and D.~Kutasov,
``Non-perturbative effects in matrix models and D-branes,''
arXiv:hep-th/0306177.
%%CITATION = HEP-TH 0306177;%%
}

\nref\TakayanagiSM{
T.~Takayanagi and N.~Toumbas,
``A matrix model dual of type 0B string theory in two dimensions,''
JHEP {\bf 0307}, 064 (2003)
[arXiv:hep-th/0307083].
%%CITATION = HEP-TH 0307083;%%
}

\nref\DouglasUP{
M.~R.~Douglas, I.~R.~Klebanov,
D.~Kutasov, J.~Maldacena, E.~Martinec and N.~Seiberg,
``A new hat for the $c = 1$ matrix model,''
arXiv:hep-th/0307195.
%%CITATION = HEP-TH 0307195;%%
}

\nref\GaiottoYF{
D.~Gaiotto, N.~Itzhaki and L.~Rastelli,
``On the BCFT description of holes in the c = 1 matrix model,''
arXiv:hep-th/0307221.
%%CITATION = HEP-TH 0307221;%%
}

\nref\GutperleIJ{
M.~Gutperle and P.~Kraus,
``D-brane dynamics in the c = 1 matrix model,''
arXiv:hep-th/0308047.
%%CITATION = HEP-TH 0308047;%%
}

\nref\SenIV{
A.~Sen,
``Open-closed duality: Lessons from matrix model,''
arXiv:hep-th/0308068.
%%CITATION = HEP-TH 0308068;%%
}

\nref\McGreevyDN{
J.~McGreevy, S.~Murthy and H.~Verlinde,
``Two-dimensional superstrings and the supersymmetric matrix model,''
arXiv:hep-th/0308105.
%%CITATION = HEP-TH 0308105;%%
}

\nref\KapustinHI{
A.~Kapustin,
``Noncritical superstrings in a Ramond-Ramond background,''
arXiv:hep-th/0308119.
%%CITATION = HEP-TH 0308119;%%
}

\nref\GiveonWN{
A.~Giveon, A.~Konechny, A.~Pakman and A.~Sever,
``Type 0 Strings in a 2-d Black Hole,''
arXiv:hep-th/0309056.
%%CITATION = HEP-TH 0309056;%%
}

\lref\joesbook{
J.~Polchinski, {\it String Theory}, Volume 2, Cambridge
University Press (1998).
}

\lref\McGreevyTA{
J.~McGreevy, S.~Murthy and H.~Verlinde,
{\it to appear}.
}

\Title{\vbox{\baselineskip12pt
	\hbox{hep-th/0309148}
        \hbox{NSF-KITP-03-76}
	\hbox{PUPT-2096}
}}{On the $S$-matrix of Type 0 String Theory}

\centerline{Oliver DeWolfe${}^{\dagger \ddagger}$,
Radu Roiban${}^*$,
Marcus Spradlin${}^\ddagger$,  Anastasia Volovich${}^\ddagger$ and
Johannes Walcher${}^\ddagger$}

\bigskip
\bigskip

\centerline{${}^\dagger$Department of Physics, Princeton University}
\centerline{Princeton, NJ 08544 USA}

\smallskip

\centerline{${}^\ddagger$Kavli Institute for Theoretical Physics}
\centerline{Santa Barbara, CA 93106 USA}

\smallskip

\centerline{${}^*$Department of Physics, University of California}
\centerline{Santa Barbara, CA 93106 USA}

\bigskip
\bigskip

\centerline{\bf Abstract}

\bigskip
The recent discovery of non-perturbatively stable two-dimensional string
backgrounds and their dual matix models allows the study of complete
scattering
matrices in string theory.
In this note we adapt work of Moore, Plesser, and Ramgoolam on the
bosonic 
string to compute the exact $S$-matrices of 0A and 0B string theory in two
dimensions.
Unitarity of the 0B theory requires the inclusion of massless soliton
sectors
carrying RR scalar charge as asymptotic states.
We propose a regularization of IR divergences and find
transition 
probabilities that distinguish the otherwise energetically
degenerate soliton sectors.
Unstable D-branes can decay into distinct soliton sectors.

\bigskip
\bigskip
\medskip

\Date{September 2003}

\listtoc
\writetoc

\newsec{Introduction}

Matrix model duals of string theory in low dimensions
were extensively studied a decade ago
(see \refs{\KlebanovQA, \GinspargIS, \PolchinskiMB} and
references therein).   According to the $c=1$ version of
this duality, strings in two dimensions
are equivalent to free fermions in an inverted
harmonic oscillator potential.  In
particular,
the linear dilaton `vacuum' of
bosonic string theory with asymptotic string
coupling $g$ corresponds, in the dual picture, to
a Fermi sea which is filled on one side of the potential up to
a distance $\mu \sim (g \sqrt{\alpha'})^{-1}$ from the top (see Fig.~1).

The most important lesson learned from the study of these matrix models
has to do with
the strength of non-perturbative effects in string theory.
The matrix model has obvious non-perturbative effects of
order $e^{-{\cal{O}}(\sqrt{\alpha'} \mu)}
\sim e^{-{\cal{O}}(1/g)}$ due to the tunneling of a single fermion
through the potential barrier to the other side.
Effects of this order, as opposed to the smaller
$e^{-{\cal{O}}(1/g^2)}$ effects
familiar from gauge theories, were found to be a general
feature of string theory even in higher dimensions \ShenkerUF,
appearing naturally when worldsheets with boundaries are included
\PolchinskiFQ.

Although these developments helped lay the groundwork for
the second superstring revolution, the discovery of D-branes,
and ultimately 
the gauge theory/gravity dualities exemplified by the AdS/CFT correspondence,
the matrix model itself went into
hibernation for a good reason:  the bosonic model is
obviously unstable.  Although there is no tachyon
in its perturbative spectrum, the theory of the half-filled
Fermi sea is non-perturbatively unstable against
tunneling through the barrier to fill up the other half of the
sea.

Now ten years later, our growing comfort with tachyon
condensation in string theory has led, following
\McGreevyKB, to a $c=1$ matrix
model renaissance
(some related work includes \refs{\MartinecKA-\GiveonWN}).
In particular, the endpoint
of the aforementioned instability, in which both
sides of the Fermi sea are filled up to the same level,
has been identified\foot{Significant progress towards  
this discovery was made in \DharGW, although that
work fell short of identifying antisymmetric
displacements of the Fermi sea as the fluctuating RR scalar of
the 0B string theory.}
as the unique stable vacuum of two-dimensional type 0B string theory
\refs{\TakayanagiSM,\DouglasUP},
 with the matrix model
reinterpreted as the dual open string dynamics of a 
collection of unstable D-particles \McGreevyKB. 
This remarkable proposal gives us a
simple and useful calculational framework for studying a complete,
non-perturbatively well-defined string theory.

\fig{Matrix model description of the
linear dilaton `vacuum' of two-dimensional
bosonic string theory (left), and the stable vacuum
of the two-dimensional 0B string theory (right).}{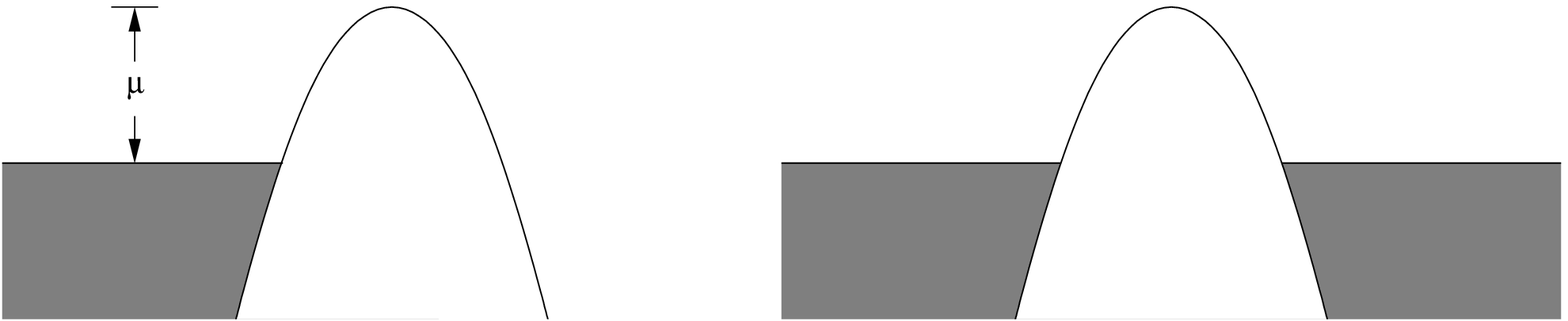}{11.0cm}

The exact spacetime solution corresponding to the double-filled Fermi
sea is not known; indeed, even the action for which it is a solution is
unknown\foot{The linear dilaton solution given in \DouglasUP\ (2.3) is
not a solution of the equations of motion coming from the effective 
action (2.4) for any functions $f_i(T)$.  Neglected terms with higher
derivatives in $T$ will contribute at the same order due to
the form of the tachyon profile.}. 
Qualitatively it resembles the linear
dilaton solution of the bosonic theory in that 
there is an asymptotically flat region in one direction where the
coupling vanishes, while the strong-coupling region in the other
direction is cut off by a ``Liouville wall,'' a condensation of the
massless ``tachyon'' field $T$ with strength proportional to the
height $\mu$ of the Fermi sea.  The 0B string theory has two massless
fields in its perturbative spectrum: the $T$ field from the NSNS
sector, and the RR
pseudoscalar $C$.  The observable of this theory is
the string theory $S$-matrix $S_{\rm ST}$ describing scattering of $T$
and $C$ quanta off of the Liouville wall.

The purpose of this paper is to study this $S$-matrix of two-dimensional
superstrings.
$S$-matrices for two dimensional string theory in various backgrounds
were also extensively studied in the previous incarnation of $c=1$
matrix models.  
The work of Moore, Plesser and Ramgoolam (MPR)
\MooreZV\ determined the exact $S$-matrix for several variations on the
inverted harmonic oscillator potential. One of these, which
they called `theory
II', had a Fermi sea filled on both sides.\foot{The `type I' family of
theories had an infinite potential wall added to prevent barrier
penetration; from a modern perspective, this somewhat {\it ad hoc}
prescription is less interesting.}
The $S$-matrix of MPR's theory II is essentially that of the 0B vacuum
except for two differences, one minor and one major.  The minor
difference is that MPR interpreted fluctuations of the left and right
Fermi seas as spacetime tachyons propagating in separate universes
(speculatively related, for example, to the inside and outside of a
black hole).  In the 0B case we instead relate left and right
fluctuations of the Fermi sea to the two separate spacetime fields
\refs{\TakayanagiSM,\DouglasUP}
\eqn\tlr{
T_{\rm L,R} = {1 \over \sqrt{2}}(T \pm C).
}
The major difference is that MPR found their $S$-matrix
to be non-unitary in theory II,
owing to the fact that their Hilbert space of asymptotic
states did not included
``soliton" sectors
corresponding to single-fermion tunneling.
We incorporate these explicitly, and find a
unitary $S$-matrix.

There is also a type 0A theory in two dimensions.  Unlike the 0B case,
which has a unique vacuum, the 0A theory has a family of vacua labeled
by an integer $q$ which measures the D0-brane charge.  There is only a
single dynamical `tachyon' field $T$ in spacetime.  The matrix model
which has recently been proposed \DouglasUP\ as dual to the 0A theory
was actually solved over a decade ago in a completely different guise
\DemeterfiCM.  Its $S$-matrices have fewer subtleties than that of 0B,
so we will start by writing them down as a useful warm-up exercise.

The plan of this paper is as follows.
In section 2 we review the central result of \MooreZV, that the
spacetime $S$-matrix is essentially the matrix model $S$-matrix convolved
with spacetime bosonization.   This result applies readily to the
0A vacua.  In section 3 we turn our attention to 0B and start
by reviewing the apparent non-unitarity
of the $S$-matrix of theory II, which arises from using an incomplete
basis of in- and out- scattering states.  Strict interpretation of
the bosonization map
dictates the addition of an infinite number of
`soliton' sectors labeled by an integer $k$ which is dual to the
zero mode of the RR field $C$.  In section 4 we compute a number of
simple amplitudes and determine the general structure of the $S$-matrix
between arbitrary soliton sectors.  Notably, we find that transition
amplitudes between different soliton sectors are plagued by the
usual infrared catastrophe of massless field theories.
We propose a regularization in which
the soliton sector vacua $|k\rangle$ are apparently
not indistinguishable, rendering the diagonalization of
the Hamiltonian difficult.
It would be interesting to see if there is
an alternate regularization which treats
the $k$-vacua on equal footing.

\newsec{The $S$-matrix for Two-Dimensional String Theory}

\subsec{The main formula}

In string theory we are interested in the spacetime
$S$-matrix for scattering $T$ (or, in the 0B theory, $T$ and $C$)
off of the Liouville wall.  In the matrix model we can easily
calculate the $S$-matrix for scattering free fermions off of the potential.
In \MooreZV\ it was shown that these two
$S$-matrices
are related to each other by the intuitive
formula
\eqn\MPR{
S_{\rm ST} \sim S_{\rm CF}
\equiv \imath_{f \to b} \circ S_{\rm FF} \circ \imath_{b \to f},
} where $\imath_{f \to b}$ is the bosonization map,
$\imath_{b \to f}$ is its inverse
and $S_{\rm CF}$ is known as the collective field
$S$-matrix \DasKA.  The symbol $\sim$ denotes an additional subtlety:
the bosonized matrix model fermion is not identical to a spacetime
field; rather the matrix model field and
the spacetime field are related to each other by a particular
background-dependent non-local field redefinition.  Fortunately, the
precise field redefinition appropriate for the 0A and 0B vacua is
known, and will be incorporated below.

We stress that the bosonization in \MPR\ is
happening in spacetime, rather than on the worldsheet.
Furthermore the formula \MPR\ does not imply an exact relation between
a non-relativistic fermion living in the matrix model potential
and
an object in
the spacetime string field theory.  Instead it only relates the
asymptotics of correlation functions (i.e., $S$-matrix elements) of
these two theories, and this only works because the spacetime theory
is asymptotically a conformal field theory far from the Liouville
wall \DijkgraafHK.
In \MPR\ it is understood that the fermionization map
$\imath_{b \to f}$ acts only on modes coming in from ${\cal{I}}^-$,
and the bosonization map $\imath_{f \to b}$ acts only on modes going
out to ${\cal{I}}^+$.

\subsec{The 0A $S$-matrix}

We begin by applying
the main formula \MPR\ to the
simpler case of the 0A theory.
Essentially nothing in this section is actually new, since in fact the
0A theory was completely solved (unbeknownst to them, we presume) a
decade ago by Demeterfi, Klebanov and Rodrigues \DemeterfiCM.  The
connection emerged only after the proposal of \DouglasUP\ that the 0A
string theory in the background of $q$ (stable) D0 branes is described
by a $U(N + q) \times U(N)$ matrix model.\foot{The applicability of the
DKR calculation to type 0A string theory was independently realized by
Kapustin \KapustinHI, who also proposed a novel duality exchanging the
string coupling and RR flux.}  After taking the double scaling limit
and integrating out the `angular' variables of the matrix, the
effective dynamics of the eigenvalues $\lambda_i$ reduces to that of
$N$ free fermions on the half line $\lambda \in (0,\infty)$ with the
Hamiltonian
\eqn\hamilA{
H = - {1 \over 2} {d^2 \over d\lambda^2} - {1 \over 2 \alpha^2} \lambda^2
+ {q^2 - {1 \over 4} \over 2 \lambda^2}, \qquad \alpha
\equiv \sqrt{2 \alpha'}.
}
Such `deformed' matrix models with a $1/\lambda^2$ potential
in addition to the
usual inverted harmonic oscillator term $-\lambda^2$ were  
extensively studied following \JevickiZG.

Let us introduce canonically normalized
fermionic mode operators
$b(\omega)$.
The empty Fermi sea $|0\rangle$ defined by $(H+\mu) |0\rangle$
satisfies
\eqn\fermisea{\eqalign{
b(\omega)|0\rangle &= 0, \qquad \omega > - \mu,\cr
b^\dagger(\omega)|0\rangle &= 0, \qquad \omega < - \mu.
}}
Since the fermions are free, the $S$-matrix of the matrix model
is completely determined in terms of the reflection
coefficient for the Hamiltonian \hamilA,
which was computed in \DemeterfiCM:
\eqn\RA{
R(\omega) =
\left| {q^2 - {1 \over 4} + \alpha^2 \mu^2 \over 4} \right|^{- i \alpha
\omega/2}
{\Gamma(\half + {|q| \over 2} - {i \over 2}
\alpha \mu + {i \over 2} \alpha \omega ) \over
\Gamma(\half + {|q| \over 2} + {i \over 2} \alpha \mu
 - {i \over 2} \alpha \omega)}.
}
Overall phase factors independent of the frequency
$\omega$ are irrelevant and have been dropped from \RA.
The physical content of the reflection coefficient \RA\ is that
the unitary map between the in and out Hilbert spaces is simply
\eqn\aaazzz{
b_{\rm out}(\omega) = R(\omega) b_{\rm in}(\omega) =
S b_{\rm in}(\omega) S^{-1},
}
where the $S$-matrix is
\eqn\SA{
S = \exp \left(-\int_{-\infty}^\infty d\omega\ \ln R(\omega)\ b^\dagger_{\rm
in}(\omega) b^{}_{\rm in}(\omega) \right).
}
Next we wish to use \MPR\ to
express \SA\ in a Fock basis of massless in/out bosons
rather than in/out fermions.  To this end we introduce
canonical bosonic modes
$a(\omega)$ such that
\eqn\bosone{
[ a(\omega), a(\omega') ] = \omega \delta(\omega + \omega')
}
with $a^\dagger(\omega) = a(-\omega)$ and
\eqn\bvac{
a(\omega)|0\rangle = 0, \qquad \omega > 0.
}
The fermionization map is given by
\eqn\fermionize{
\imath_{b \to f}:\qquad a(\omega) \to \int_{-\infty}^\infty
d\xi\ b(\xi - \mu) b^\dagger(\xi - \omega - \mu).
}
Note that $\mu$ enters in this formula in order to ensure that \bvac\ is
compatible with \fermisea.
Of course there are two separate copies of the fermionization
map \fermionize\ which
operate separately on the in/out Fock spaces.

\fig{Scattering experiment in the 0A theory.  Each incoming
bosonic mode in the spacetime picture fermionizes into a particle/hole
pair.  These scatter individually 
off the potential with amplitudes
\RA\ and return to infinity, where they are re-bosonized into
some configuration of outgoing tachyons in spacetime.}{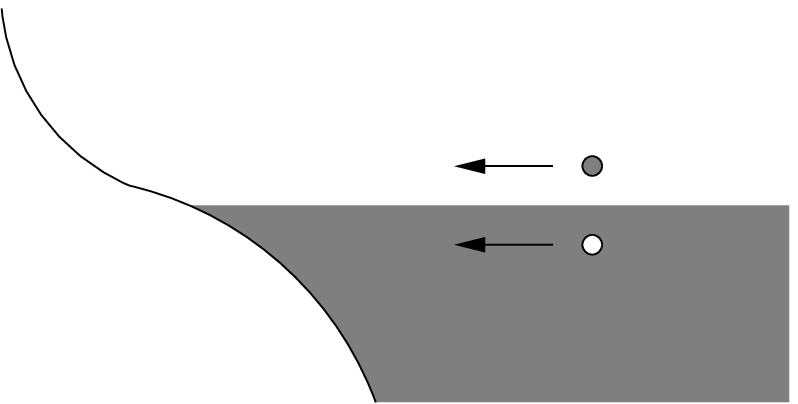}{5.0cm}

The $S$-matrix for the scattering of $m \to n$ tachyons
in the 0A theory is
simply
\eqn\SmatrixA{
S_{\rm CF}(\omega_1,\ldots,\omega_m|\omega_1',\ldots,\omega'_n) \equiv
\langle 0 | \prod_{j=1}^n a_{\rm out}(\omega_j')
 \prod_{k=1}^m a_{\rm in}(-\omega_k)|0\rangle,
}
where $a_{\rm out/in}$ are expressed in terms of $b_{\rm out/in}$
using \fermionize.
We use the convention that the arguments
$\omega$ appearing in $S(\cdots|\cdots)$ are always positive.
One can obtain explicit formulas for arbitrary amplitudes
by substituting \fermionize\ and using \aaazzz.  A set of
graphical rules for immediately writing down any amplitude
were presented in \MooreZV.  For example, the $1 \to n$
amplitude\foot{We use the standard convention that ${\cal{A}}$ is
equal to $S$ with the overall energy conservation
delta-function removed.} is given by
\eqn\oneton{
{\cal{A}}_{\rm CF}(\omega|\omega_1,\ldots,\omega_n)
=
\sum_{k=0}^{n-1} (-1)^k \sum_{ \{ j_1,\ldots,j_k \}}
\int_{\omega_{j_1}
+ \cdots + \omega_{j_k}}^\omega
\!\!\!\!\!\!\!\!\!\!\!\!\!\!\!\!\!\!\!\! d\xi \ R(\omega-\xi) R^*(-\xi)
}
where $\omega = \omega_1+\cdots+\omega_n$ and the
final sum is over all possible length $k$ subsets of
$\{1,\ldots,n\}$.

Before proceeding, let us note that  the picture shown in Fig.~2 depicts
the case $q \ne  0$ so that the potential in \hamilA\ provides
a repulsive barrier at small $\lambda$.  However for $q=0$
the potential is attractive near the origin.
Nevertheless there is no bound state and the scattering matrix
remains well-defined and unitary, at least for $\alpha \mu \ne \ha$.
For $\alpha \mu = \ha$, problems appear to arise in the the formula
\RA\ for the reflection coefficient.  We have not
investigated whether this is indicative of some
interesting new strong coupling physics
or whether it is just a coordinate singularity with no
physical effects.

\subsec{Three comments on the two-dimensional $S$-matrix}

\noindent
{\bf (1).} We have not yet taken into account the non-local transformation
alluded to beneath \MPR.  The spacetime tachyon field $T = T(t,\phi)$
is a function of time and the Liouville coordinate $\phi$, whereas
the field obtained by bosonizing the matrix model fermion is a function
of time and the eigenvalue coordinate $\lambda$.  The time coordinate
$t$ is identified between the matrix model and spacetime, but the
coordinates $\phi$ and $\lambda$ are related by a non-local transformation.
This transformation is most easily written in momentum space,
where it becomes a momentum-dependent phase factor
(the so-called `leg-pole factor')
\eqn\legA{
e^{i \delta(\omega)} \equiv
|q^2 - {\textstyle{1 \over 4}}|^{i \alpha \omega/4} {
\Gamma(-{i \over 2} \alpha \omega)
\over \Gamma({i \over 2} \alpha \omega)}.
}
The string theory $S$-matrix is related to the $S_{\rm CF}$
in \SmatrixA\ by
\eqn\aaa{
S_{\rm ST}(\omega_1,\ldots,\omega_n|\omega_1',\ldots,\omega_m')
= S_{\rm CF}(\omega_1,\ldots,\omega_n|\omega_1',\ldots,\omega_m')
\prod_{i=1}^n e^{i \delta(\omega_i)}
\prod_{j=1}^m e^{i \delta(\omega_j')}.
}

\noindent 
{\bf (2).}
The result we wrote in
\oneton\ allows for particle production, yet when written
in terms of fermions the theory is free and obviously cannot
have any particle production!
The resolution of this apparent puzzle lies in the fact
that, as emphasized by MPR, a multi-boson state
is not normal ordered with respect to the Fermi sea.
Instead, we have (schematically)
\eqn\aaa{
\eqalign{
\underbrace{a \cdots a}_n|0\rangle
&=  \underbrace{\int :\!b^\dagger b\!: \cdots \int :\!b^\dagger b\!:}_n
 |0\rangle
\cr
&=~:\!\int \underbrace{b b^\dagger \cdots b b^\dagger}_n \!:|0\rangle
+ :\!\int \underbrace{b b^\dagger \cdots b b^\dagger}_{n-1}
\!:|0\rangle
+ \cdots +~:\!\int b b^\dagger 
\!:|0\rangle.
}}
Even though fermion number is strictly conserved, bosonic
particle production can occur because the bosonization map
relates a state with $n$ bosons to a state of indefinite fermion
number.

\noindent
{\bf (3).}  Finally, and most importantly, this $S$-matrix
has no right to exist!
We claim to have calculated an $S$-matrix for massless particles
in two dimensions, but such an object suffers from two well-known
pathologies.  The first is that
the $S$-matrix is generally ill-defined in the presence of massless
particles
due to infrared
divergences associated with the proliferation of soft quanta.
Consider the unitarity equation in the $1 \to 1$ channel:
\eqn\unitarity{
1 = {1 \over \omega^2} | {\cal{A}}(\omega|\omega)|^2 +
{1 \over \omega} \sum_{n = 2}^\infty
\prod_{i=1}^n\int_0^\infty {d \omega_i \over \omega_i}
\delta \left( \sum_{i=1}^n \omega_i - \omega\right) | {\cal{A}}(\omega|
\omega_1,\ldots,\omega_n)|^2.
}
In a typical massless theory one needs to introduce
an IR cutoff $\Lambda$ in order to render $S$-matrix elements
well-defined.  Then the amplitude to emit any fixed number $n$
of quanta with $\omega_i \approx 0$ vanishes linearly as $\Lambda \to 0$.
However, these amplitudes cannot be dropped without violating
unitarity because they contribute a finite amount to \unitarity\ after
performing the
$\omega$ integrals and summing over the number $n$ of emitted soft photons.
Essentially, an extra factor of $\Lambda^{-2}$ comes from regulating the
divergences in the density of states factor
\eqn\aaa{
\sum \prod \int_0^\infty { d\omega_i \over \omega_i}.
}
In the present case, the amplitude to emit any fixed number
of quanta is actually finite and non-zero
without the need for an IR regulator.
This is only consistent because all amplitudes vanish linearly
as any one of the $\omega_i$ is taken to zero.  For example,
\eqn\aaa{
{\cal{A}}_{\rm CF}(\omega|\omega) = 
\int_0^\omega d\xi\ R(\omega-\xi) R^*(-\xi)
= \omega + {\cal{O}}(\omega^2).
}
This fact renders all of the $\int {d \omega \over \omega}$ integrals
in \unitarity\ finite, so there is a finite contribution to the
probability for any given number $n$ of emitted quanta.

This infrared catastrophe problem is a general feature of massless
$S$-matrices in any dimension.    In two dimensions there is yet
another problem.  Because of the special kinematics of massless
particles in two dimensions, we cannot expect to tell
the difference between, for example, a right-moving boson of
energy $\omega$ and two right-moving bosons of energy $\omega/2$.
This problem is solved in the present case by the fact that
the $S$-matrix has a huge symmetry group related to $W_\infty$
which is manifest in the fermionic formulation but obscure
in the bosonic picture.
One of the consequences of this large symmetry group is that
it does allow us to distinguish these states \GinspargIS.

\newsec{The 0B $S$-matrix}

Let us start by constructing the free fermion $S$-matrix $S_{\rm FF}$
for the double-filled Fermi sea.  This essentially is the $S$-matrix for
the so-called `theory II' of Moore, Plesser and Ramgoolam \MooreZV.
The seeming non-unitarity of the resulting spacetime $S$-matrix $S_{\rm
CF}$ will then be corrected by properly including the soliton sectors
$|k\rangle$ implied by bosonization.

\subsec{The free fermion $S$-matrix}

Type 0B string theory in two dimensions has two physical
fields in its perturbative spectrum:  the spacetime `tachyon' $T$
(which is actually massless) and a massless RR
scalar $C$.  According to \refs{\TakayanagiSM,\DouglasUP} the
dual matrix model involves free fermions with the familiar
inverted harmonic oscillator Hamiltonian
\eqn\hamilB{
H = - \ha {d^2 \over d\lambda^2} - {1 \over 2 \alpha^2} \lambda^2 \,,
}
and both sides of the Fermi sea filled.
Fluctuations of $T_{\rm L,R} \equiv {1 \over \sqrt{2}}(T \pm C)$
are identified with fluctuations of the left and right Fermi seas.
Clearly $T_{\rm L}$ and $T_{\rm R}$ are perturbatively decoupled
and only interact with each other by non-perturbative tunneling
of fermions through the barrier.

Let us start by writing down the free fermion $S$-matrix.
At this stage the only new ingredient as compared to the previous
0A discussion is that we now have twice as many fermionic modes.
The second quantized fermion field $\hat{\psi}(t,\lambda)$ is
now expanded as
\eqn\aaa{
\hat{\psi}(t,\lambda) = \int_{-\infty}^\infty d\omega\ e^{i \omega t}
\left[
b_+(\omega) \psi^+(\omega, \lambda) + b_-(\omega) \psi^-(\omega, \lambda)
\right],
}
where $\psi^\pm(\omega,\lambda)$ are a complete set of
normalized even/odd
wavefunctions of the Hamiltonian \hamilB\ and the mode operators satisfy
\eqn\aaa{
\{ b^{}_\epsilon(\omega), b^\dagger_\eta(\omega') \} = \delta_{\epsilon
\eta} \delta(\omega-\omega'), \qquad \epsilon,\eta \in \pm.
}
We will pass back and forth between this basis and
\eqn\caution{
b_{\rm L,R} = {1 \over \sqrt{2}} (b_+ \pm b_-).
}
The empty Fermi sea is defined by \fermisea\ separately for
both sets of oscillators.

A single excitation of the Fermi sea sent in from the left or
right can either scatter off the barrier or tunnel through it.
The transmission and reflection coefficients were
calculated in \MooreZV:
\eqn\rtB{
R(\omega) =  {1 \over \sqrt{2 \pi}}
(i \alpha \mu)^{i \alpha \omega} 
\Gamma(\half
+ i \alpha \mu
- i \alpha\omega), 
\qquad
T(\omega) =-i e^{\pi \alpha (\omega - \mu)} R(\omega).
}
Note that
\eqn\randtsq{
|R(\omega)|^2 = {1 \over 1 + e^{-2 \pi \alpha (\mu - \omega)}}, \qquad
|T(\omega)|^2 = {1 \over 1 + e^{+2 \pi \alpha (\omega - \mu)}}
}
and
\eqn\unitary{
|R|^2 + |T|^2 = 1, \qquad R T^* + T R^* = 0.
}
The in and out Hilbert spaces are related by
the unitary transformation
\eqn\trans{
\pmatrix{ b^{\rm out}_{\rm L} \cr
b^{\rm out}_{\rm R}} =
\pmatrix{ R& T \cr
T & R } \pmatrix{ b^{\rm in}_{\rm L} \cr
b^{\rm in}_{\rm R}}.
}
This transformation becomes diagonal
in the $\pm$ basis,
\eqn\aaa{
b^{\rm out}_\pm(\omega) = R^\pm(\omega) b^{\rm in}_\pm(\omega),
}
where
\eqn\aaa{
R^\pm(\omega) = R(\omega) \pm T(\omega)
}
We conclude that the free-fermion $S$ matrix
exactly factorizes into
two commuting pieces:
\eqn\SB{
S = S_+ S_-,
\qquad
S_\pm = \exp \left(-\int_{-\infty}^\infty
d\omega\ \ln R^\pm(\omega)\  b_{\pm {\rm in}}^\dagger(\omega)
b^{}_{\pm {\rm in}}(\omega)\right).
}
In the 0B theory we combine the two spacetime fields $T$, $C$
into $T_{\rm L,R}$, which are separately
fermionized into $b_{\rm L,R}$.
We caution the reader against concluding from \tlr\ and \caution\ that
the factorization of the fermionic $S$-matrix into $S = S_+ S_-$ implies
that the bosonic $S$-matrix
factorizes into $S = S_T S_C$.  This factorization would imply that the
$T$ and $C$ fields do not talk to each other! 
This does
not happen
because the
modes $b_\pm$ are not simply related to those of $T$ and $C$ because
bosonization is non-linear:
\eqn\aaa{\eqalign{
{1 \over \sqrt{2}}(a_{\rm L}(\omega) \pm a_{\rm R}(\omega))
&= {1 \over \sqrt{2}} \int_{-\infty}^\infty d\xi \left[ b_{\rm L}(\xi
- \mu) b^\dagger_{\rm L} (\xi - \omega - \mu)  \pm 
({\rm L} \leftrightarrow {\rm R})\right]
\cr
&\ne \int_{-\infty}^\infty\ d\xi\ b_\pm(\xi - \mu)
b_\pm^\dagger(\xi-\omega-\mu).
}}

\subsec{The problem with `Theory II'}

Let us consider a scattering experiment where we prepare a single
quantum of the $T_{\rm L}$ field and send it in towards the
Liouville wall.  This bosonizes into the particle/hole pair
shown in Fig.~3.

\fig{This figure shows the matrix model description of the process
where we send in a single quantum of $T_{\rm L}$ from infinity towards
the Liouville wall.}
{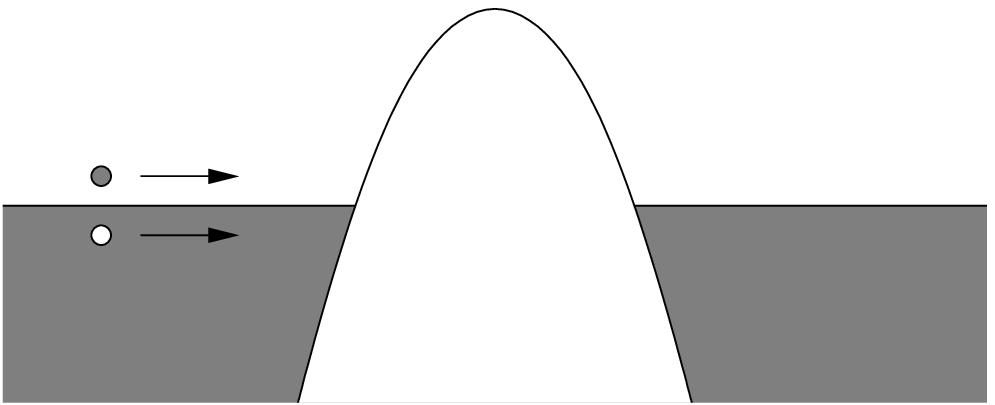}{5.5cm}

Since the fermions are free and the total fermion number is conserved,
there are obviously exactly four possible final states for this
process, shown in Fig.~4.
We have labeled the final states in order of decreasing probability.
Note that it is always easier for a particle to tunnel through the
barrier than it is for a hole to tunnel, since
the barrier is always narrower for the particle.
When $\mu$ is large, (b) and (c) are suppressed relative to
(a) by a factor of $e^{-\pi \alpha \mu}$
and (d) is suppressed by a factor
of $e^{-2 \pi \alpha \mu}$.

\fig{The four possible final states of the experiment depicted
in Fig.~3, labeled (a), (b), (c), and (d)
in order of decreasing probability.}
{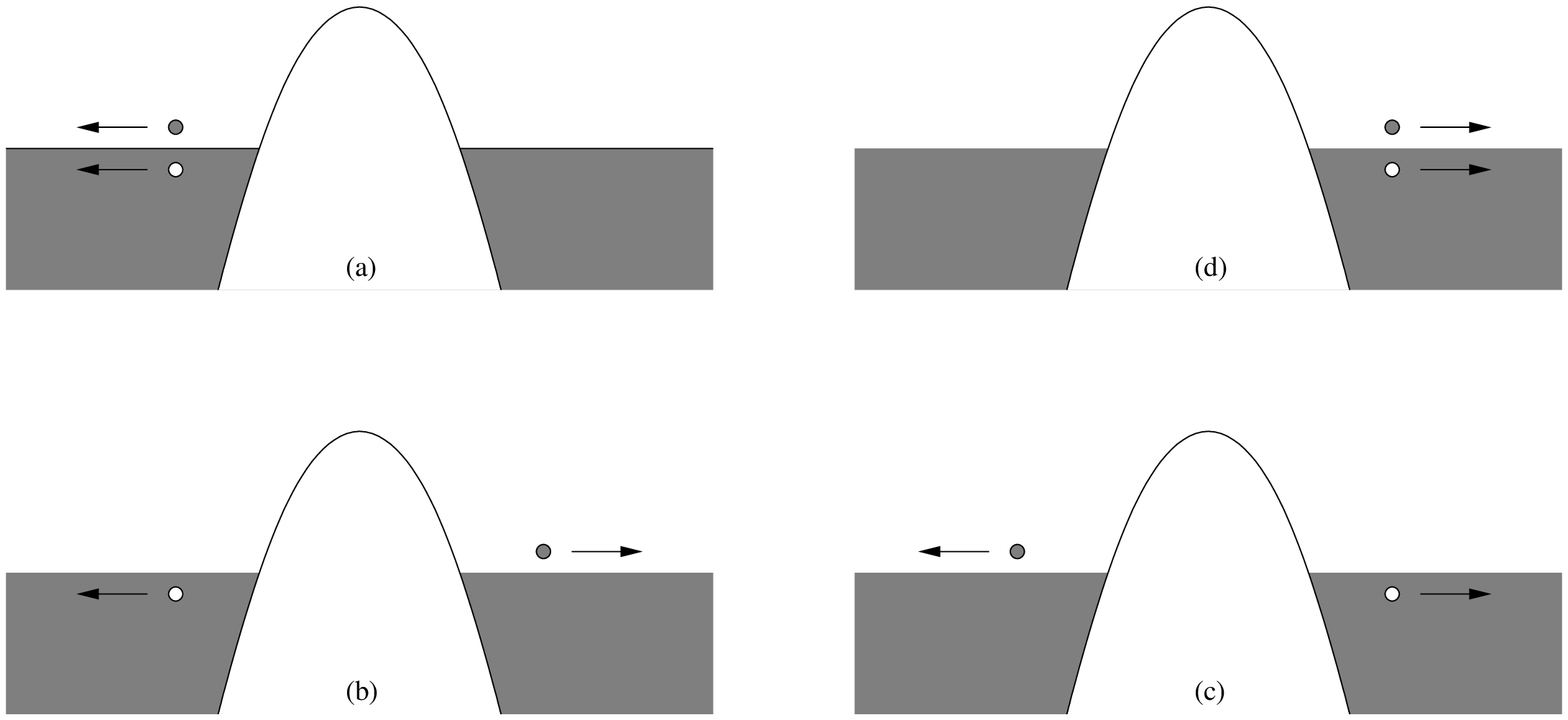}{12.0cm}

The spacetime interpretation of (a) and (d) is clear:  they
correspond respectively to a single quantum of $T_{\rm L}$
or $T_{\rm R}$ emerging from the
Liouville wall and going out to infinity.
The spacetime interpretation of (b) and (c) is more subtle.
In fact these processes were not given any spacetime interpretation
in \MooreZV, and as a result their $S$-matrix was considered not unitary.

The philosophy of this paper is that the fundamental formula
\MPR\ should be taken as literally true.
Since the free fermion $S$-matrix $S_{\rm FF}$ is manifestly
unitary, the failure of $S_{\rm ST}$
to be unitarity must necessarily be due to a failure in
the bosonization map used in the previous subsection.
This failure is nothing more than the familiar fact that 
finite energy excitations of $T$ and $C$ on the vacuum $|0\rangle$
do not give a complete basis for the image of the bosonization map.
Of course the $S$-matrix cannot be unitary if an incomplete
basis of in- and out- scattering states is used \CallanUB.
In the next subsection we review the construction of the
soliton sectors necessary to provide a complete basis.

\subsec{Bosonization}

We follow the conventions of \joesbook.
Consider the holomorphic part of a compact
boson $X$ given by the mode expansion
\eqn\partialx{
\partial X(z) = - i \sum_m {a_m \over z^{m + 1}}, \qquad
a_m = {1 \over 2 \pi} \oint dz\ z^m \partial X(z).
}
The modes satisfy
\eqn\aaa{
[ a_m, a_n] = m \delta_{m,-n}.
}
The Hilbert space 
splits into disconnected sectors
${\cal{H}}_k$,
where $k$ labels the zero mode momentum and
each ${\cal{H}}_k$ is a Fock space
built on the state $|k\rangle$ satisfying
\eqn\aagh{
a_m |n \rangle = 0 \qquad \forall \; m>0, \qquad \qquad a_0 |k \rangle
= k |k\rangle.
}
Bosonization relates $\partial X$ to a complex fermion $\psi$ according
to
\eqn\aaa{
i \partial X =~:\!\psi \psi^\dagger\!:.
}
If we write $\psi$ as
\eqn\aaa{
\psi(z) = \sum_{r \in \IZ + \ha} z^{-r-1/2} b_r,
}
where
\eqn\aaa{
\{ b_r, b_s^\dagger \} = \delta_{rs}
}
and the vacuum $|0\rangle$ satisfies
\eqn\aaa{
b_r|0\rangle = 0 \qquad \forall \; r > 0, \qquad \qquad b_s^\dagger|0\rangle = 0
\qquad \forall \; s < 0,
}
then \aagh\ implies that
\eqn\aaa{
a_n = \sum_{r \in \IZ + \ha} ~:\!b_r b^\dagger_{r - n}\!:.
}
In particular, for $n = 0$ we find
\eqn\aaa{
a_0 = \sum_{r \in \IZ + \ha} ~:\!b_r b^\dagger_{r}\!: =
\sum_{r < 0} b_r b^\dagger_{r} - \sum_{r > 0} b_r^\dagger b_r,
}
which is just the fermion number operator ($\# {\rm particles} -
\# {\rm holes}$).

The Hamiltonian is
\eqn\aaa{
L_0 = {1 \over 2} \, a_0^2 + \sum_{n > 0} a_{-n} a_n =
\sum_{r > 0} r ( b_{-r} b_{-r}^\dagger - b_r^\dagger b_r).
}
The state $|k\rangle$ may be identified in the fermionic
picture as the unique state with fermion number $a_0 |k\rangle = k |k\rangle$
and energy $L_0|k\rangle = {k^2 \over 2} |k\rangle$:
\eqn\kone{
\eqalign{
|k\rangle &= b_{-k + \ha} \cdots b_{-\ha} |0\rangle, \qquad k > 0,\cr
|k\rangle &= b_{-k - \ha} \cdots b_{\ha}^\dagger |0\rangle, \qquad k < 0.
}}
When the holomorphic and antiholomorphic pieces of $X(z)$ are combined,
$|k\rangle$ can be written in the familiar way in the bosonized
theory as a soliton,
\eqn\aaa{
|k\rangle = e^{i k X}(0)|0\rangle.
}
For this reason we will sometimes refer to $|k\rangle$ as the
charge-$k$ soliton.

It is useful to consider an operator ${\cal{O}}$ which satisfies
\eqn\odefone{
[ a_m, {\cal{O}} ] = \delta_{m,0} {\cal{O}}.
}
It follows that ${\cal{O}}$ is the raising operator conjugate to
$a_0$, in that
\eqn\odeftwo{
{\cal{O}}|k\rangle = |k{+}1\rangle.
}
It is easy to work out how the action ${\cal{O}}$ is represented
on the fermionic modes,
\eqn\odefthree{
{\cal{O}} b_r = b_{r - 1} {\cal{O}}.
}
This operator is of course nothing other than the spectral flow
operator \joesbook\ (not from the NS to R sector, but from NS past
R back to NS again).  We remind the reader again that this
spectral flow operator is acting in spacetime, not on the worldsheet.

The crucial lesson is that the full fermionic Hilbert space
built by acting with the creation operators $b_r$ and $b_s^\dagger$,
for $r < 0$ and $s>0$, is equivalent not to a single bosonic Fock
space, but to an infinite family of Fock spaces.  Each bosonic Fock
space has a ground state $|k\rangle$ where $k$ is dual to the
net fermion number.
Bosonization is a unitary map
\eqn\aaa{
\imath_{f \to b} : {\cal{H}} \to \bigoplus_{k=-\infty}^\infty
{\cal{H}}_k.
}

When we wrote down the 0A $S$-matrices in the previous section, however,
we did not worry about the different ${\cal{H}}_k$ sectors.
The reason is that
fermion number is strictly conserved in the matrix model, so it is
perfectly consistent to work in a superselection sector with fixed
fermion number (equal to zero, if we normal order with respect to the
Fermi sea).

In the 0B theory, on the other hand, we have two fermions $\psi_{\rm
L,R}$ which bosonize separately into $T_{\rm L,R}$.  Therefore the
vacuum state should be labeled by two zero-mode indices $|k_{\rm L},
k_{\rm R}\rangle$.  The total fermion number $k_{\rm L} + k_{\rm R}$
is still exactly conserved, so just as in the 0A case we can
consistently work in a superselection sector with fixed $k_{\rm L} +
k_{\rm R} = 0$.  However, the relative fermion number $k \equiv {1
\over 2} (k_{\rm L} - k_{\rm R})$ is not conserved and is in general
changed when fermions tunnel through the potential.
Note that $k$ is related to the zero mode of the $C$ collective
field\foot{The notation $C_0$ means `the zero mode of $\partial C$',
in the same sense that $a_0$ is the zero mode of $\partial X$ in \partialx.
We point this out only to highlight that it is {\it not}
the zero mode of $C$ itself.}:
\eqn\aaa{
C_0 |k \rangle = {1 \over \sqrt{2}} ( T_{{\rm L}, 0}  - T_{{\rm R},0})
|k\rangle
= {1 \over \sqrt{2}}(k_{\rm L} - k_{\rm R})|k\rangle
= \sqrt{2} k |k\rangle.
}
It follows from \kone\ that the ground state in the charge $k$ sector
is
\eqn\kvac{
|k\rangle = b_{{\rm L}, -k + \ha} \cdots b_{ {\rm L}, - \ha}
b_{{\rm R}, k - \ha}^\dagger \cdots b_{{\rm R}, \ha}^\dagger |0\rangle
= e^{i \sqrt{2} k C}(0)|0\rangle.
}
As written this formula is for $k > 0$; for $k < 0$ one
interchanges L and R.

In summary, the $S$-matrix of theory II failed to be unitary because
the basis of asymptotic scattering states used in the calculation
was not complete.  The complete Hilbert space is $\bigoplus {\cal{H}}_k$,
and unitarity is manifest if we allow states built on arbitrary $|k\rangle$
to appear in the $S$-matrix.

\subsec{Discretization and the definition of the $k$-vacua}

In the previous subsection we used fields with discrete mode expansions,
rather than the continuous mode expansions we used in
section 2 for the 0A theory.
We will see below that the 0B theory,
unlike 0A, unfortunately does suffer from the
infrared problems one usually encounters in massless theories.
The essence of the problem
is that the solitons $|k\rangle$ 
are coherent states built out of quanta with zero energy.
In order to tame the infrared catastrophe, we will follow \CallanUB\ and 
introduce an infrared
cutoff by making time periodic with period $2 \pi/\beta$.  This
cutoff restricts all energies to be multiples of $\beta$.  States then
carry discrete indices as in the previous subsection, according
to the dictionary
\eqn\somemodes{
b_r \leftrightarrow b(r \beta), \qquad a_n \leftrightarrow
a(n \beta).
}
At
the end of any calculation involving some
modes of the form \somemodes,
we take $\beta \to 0, r,n \to \infty$ keeping $\beta r$,
$\beta n$ fixed.
Note that we put antiperiodic boundary conditions on the
Fermi fields in compactified time.

The exception to the latter scaling is that the index $k$
of the ground states
$|k\rangle$ is held fixed in the continuum limit, so
\eqn\kvacbeta{
|k\rangle = \lim_{\beta \to 0}
b_{\rm L}((-k + \half) \beta)
\cdots
b_{\rm L}(- \half \beta) b_{\rm R}^\dagger((k-\half)\beta) \cdots
b_{\rm R}^\dagger(\half \beta)|0\rangle.
}
Therefore the states $|k\rangle$ become degenerate (at zero energy)
in the continuum
limit, so we will refer to them as the $k$-vacua.
Intuitively, $|k\rangle$ may be thought of as the state where
$k$ particles are skimmed from the surface of the right
Fermi sea and gently lowered onto the surface of the left Fermi sea.

This intuitive picture is made concrete by generalizing
the operator ${\cal{O}}$ defined in \odefone, \odeftwo\ and \odefthree\ to
the present case of the double-filled Fermi sea.
Essentially we now want ${\cal{O}} = {\cal{O}}_{\rm L}
{\cal{O}}_{\rm R}^\dagger$, which satisfies
\eqn\aaa{\eqalign{
{\cal{O}} |k\rangle &= |k{+}1\rangle,\cr
[ a_{{\rm L}, m}, {\cal{O}} ] = \delta_{m, 0} {\cal{O}}, &\qquad
[ a_{{\rm R}, m}, {\cal{O}} ] = -\delta_{m, 0} {\cal{O}},\cr
{\cal{O}} b_{{\rm L}, r } = b_{{\rm L}, r - \beta} {\cal{O}},
&\qquad
{\cal{O}} b_{{\rm R}, r } = b_{{\rm R}, r+ \beta} {\cal{O}}.
}}
The last  line makes manifest the fact that the raising operator
${\cal{O}}$ amongst the $k$-vacua works in the continuum
limit $\beta \to 0$, $\beta r$ fixed
by raising the left Fermi sea and lowering the right Fermi
sea by infinitesimal amounts (see Fig.~5).

These states may seem indistinguishable,
in the continuum limit, from the `bare' Fermi sea
$|0\rangle$, but it is essential to keep track of them in order to get
a unitary $S$-matrix.  In particular, the configurations in Fig.~4(b)
and (c) are orthogonal to all states in the Fock space built on
$|0\rangle$!

\fig{To construct the $|k\rangle$ vacuum one skims $k$ particles
from top of
the right Fermi sea and deposits them on top of the left Fermi sea
(or vice versa, for $k < 0$).}{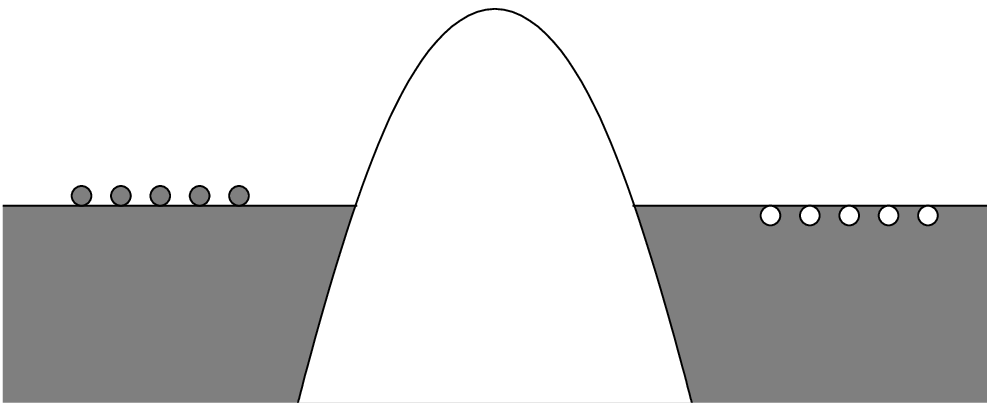}{5.5cm}

\subsec{The $k$-vacua in spacetime: D-instanton charge}

We have claimed that the spectrum of scattering states in 0B string theory is incomplete without including the ``solitonic" ${\cal{H}}_k$ sectors.  Here we discuss further the spacetime interpretation of these sectors.  Given that the $k$-changing transitions in the matrix model correspond to eigenvalue tunneling, which as reviewed in the introduction is a process of magnitude $e^{-{\cal O}(1/g)}$ in spacetime, it should come as no surprise that the ${\cal{H}}_k$ sectors are related to the physics of D-instantons.

The effective action for the 0B fields possesses, to all orders in string perturbation theory, a shift symmetry in the RR scalar $C$ \DouglasUP,
\eqn\aaa{
{\cal T}_a: \quad C \rightarrow C + a \,.}
Such a symmetry, or its higher-rank $p$-form generalization, is generic for Ramond-Ramond fields.  It implies the existence of a Noether current $J_\mu$, and as is the case in ten-dimensional IIB string theory, the objects possessing the RR scalar charge are the D-instantons.  The D-instanton number $n$ in a spacetime process can be evaluated by integrating over a circle at infinity,
 \eqn\RRcharge{n =  \int_{S^1} dx^\mu J_\mu \,.}
For example, in the expression given for the leading-order 0B effective action in \DouglasUP, the RR scalar appears as
\eqn\aaa{
S_{{\rm 0B}} \supset - {1 \over 8 \pi} \int d^2x \sqrt{-g} f_3(T) (\nabla C)^2 \,,
}
where $f_3(T)$ is a function of the tachyon field identified there as $f_3(T) = e^{-2T}$.
The Noether current is then
\eqn\DougNoether{J_\mu = f_3(T) \partial_\mu C \,,}
where we assumed a flat background metric, and \RRcharge\ by Stokes theorem is equivalently expressed as an integral over all spacetime of a density,
\eqn\PonDens{n = \int dJ = \int f_3'(T) \, dT \wedge dC \,.
}
This expression is the 2D analog of the Pontryagin density $\int F \wedge F$
for 4D Yang-Mills instantons.

Consider now the symmetry ${\cal T}_a$ acting on the $k$-vacua.  Using \kvac , we find\foot{We have not forgotten the nonlocal transformation advertised beneath \MPR\ between the collective field used to define \kvac\ and the spacetime field of 0B string theory.  However, it can be ignored for the zero mode of the field, since as discussed in the next section, the appropriate leg pole factor is unity at zero energy.}
\eqn\kcharge{{\cal T}_a: \quad |k \rangle \rightarrow e^{i\sqrt{2}ka} |k \rangle \,.
}
States in the matrix model with nonvanishing difference between left and right fermion numbers (relative to the undisturbed Fermi seas) are therefore charged under ${\cal T}_a$.  The existence of the $k$-vacua breaks the shift symmetry of $C$ non-perturbatively, though a residual identification renders the field compact:
\eqn\aaa{C \sim C + \sqrt{2} \pi \,.
}
The transformation \kcharge\ indicates that the $k$-vacua have nonzero RR charge.  Hence for any state in the sector ${\cal H}_k$, we must have
\eqn\RRstate{\int d \phi  \, J_0 = k
}
at some moment in time, where we remind the reader
that $\phi$ is the spacelike Liouville coordinate.
A D-instanton with charge $n$ \RRcharge\ can then mediate classically forbidden transitions between a state with charge $k$ and one of charge $n+k$, in accordance with our intuition based on the $e^{-{\cal O}(1/g)}$ strength of single-eigenvalue tunneling.  The $k$-vacua, then, have a relationship to D-instantons precisely analogous to that in Yang-Mills theory of the ``winding vacua" to Yang-Mills instantons.

In the matrix model description, we found that the $k$-vacua are degenerate in energy.
This is again closely analogous to the Yang-Mills case.  In order to dominate the path integral, instantons must have finite action.  A consequence of this is that the action must vanish at infinity where the charge \RRcharge\ is measured; in Yang-Mills theory, this requires $F_{\mu\nu} = 0$ asymptotically.  But the integral defining the charge $k$ of the state \RRstate\ is identical to that of the fields at infinity around a charge-$k$ instanton\foot{This is true when the fields in \RRstate\ approach the same value at infinity, effectively compactifying the spacelike slice.}.  In Yang-Mills theory, this implies the existence of a state of each charge $k$ with vanishing field strength and hence zero energy.  Although we lack an adequate effective action for the spacetime fields, it seems probable that the finiteness of the action for configurations of $k$ D-instantons also implies the existence of zero energy states of charge $k$, namely the $k$-vacua. 

It would be highly interesting to find the correct effective action for 0B string theory and construct a D-instanton solution.  If the correct action is some iteration of (2.4) in \DouglasUP, it is clear that more than simply the $C$-field will be involved, as any nonzero value for the RR current \DougNoether\ will backreact on the other fields.

Because of the D-instanton processes, the $k$-vacua will mix, and will not be eigenvectors of the full Hamiltonian.  It is then natural to inquire what the energy eigenstates are.
In the Yang-Mills case, the winding vacua are related by a large gauge transformation and hence possess precisely equivalent physics.  The diagonalization of the Hamiltonian then proceeds analogously to the study of Bloch waves in quantum mechanics: the eigenvectors of the Hamiltonian are a band of $\theta$-vacua, and the energy density of the states goes like $\cos \theta$. Similarly, one may define for 0B string theory a set of $c$-vacua:
\eqn\CVac{
| c \rangle \equiv \sum_k e^{-i \sqrt{2} k c} |k \rangle \,.}
These states are conjugate to the $k$-vacua, and are eigenvectors of the zero mode of $C$.  $\theta$-vacua in Yang-Mills theory are superselection sectors; we expect however that in our two-dimensional case, transitions between the $c$-vacua can be effected by sending in energy in the form of $C$ or $T$ quanta.

The analogy to Yang-Mills physics persists only if the $k$-vacua prove to be indistinguishable.  We have already argued in the matrix model that the  $k$-vacua are energetically degenerate; we have not, however, shown they have equivalent physics.  In the next section, we calculate a few basic amplitudes and provide evidence that the sectors of different $k$ are, in fact, different with our choice of infrared regularization.
If this turns out to be the case, the diagonalization of the Hamiltonian will be more involved.

\newsec{Some 0B Amplitudes}

In this section we use the prescription of the previous section to
calculate several interesting amplitudes in the 0B theory.  These
examples will highlight how easy it is to calculate arbitrary
non-perturbative $S$-matrix elements as exact functions of $\mu$.
Furthermore we will verify our claim that the usual infrared
catastrophe of massless theories is present.  In general amplitudes
must be calculated in the discretized theory at $\beta > 0$, where the
theory is manifestly unitary.  In the limit $\beta \to 0$ only
probabilities are in general well-defined.

\subsec{Leg-pole factors}

Before proceeding we should note that the leg-pole factors
which relate the collective field $S$-matrix $S_{\rm CF}$ to the
string theory $S$-matrix $S_{\rm ST}$ were obtained in \DouglasUP\ and
are given by
\eqn\legB{
e^{i \delta_T(\omega)}
= {\Gamma({i \over 2} \alpha \omega) \over
\Gamma(-{i \over 2} \alpha \omega)}, \qquad
e^{i \delta_C(\omega)} = {\Gamma(\ha + {i \over 2} \alpha \omega)
\over \Gamma(\ha - {i \over 2} \alpha \omega)}
} for the NSNS and RR fields respectively.  It has been emphasized
\refs{\NatsuumeSP, \PolchinskiJP, \PolchinskiMB} that these factors
are essential for obtaining a local theory of gravity in two spacetime
dimensions.  However, since most of our calculations below involve
simple transition probabilities, we will consider the collective field
description to be adequate and neglect the leg pole factors.
Note that the leg-pole factors go to 1 at $\omega = 0$, so these factors
do not affect the zero energy particles/holes constituting the
$k$-vacua.

\subsec{Transitions between $k$-vacua}

We start with very explicit
calculations of the simplest amplitudes.  Using the definition \kvac\ we 
consider for $k_1, k_2 \ge 0$ the amplitude
\eqn\formone{\eqalign{
\langle k_1 | S |k_2 \rangle &=
\langle 0 | b^{\rm out}_{{\rm R}, \ha} \cdots b^{\rm out}_{{\rm R}, k_1 - \ha}
b^{{\rm out} \dagger}_{{\rm L}, -\ha} \cdots
b^{{\rm out} \dagger}_{{\rm L}, -k_1 + \ha}
b^{\rm in}_{{\rm L}, -k_2 + \ha} \cdots b^{\rm in}_{ {\rm L}, - \ha}
b^{{\rm 
in} \dagger}_{{\rm R}, k_2 - \ha} \cdots
b^{{\rm in} \dagger}_{{\rm R}, \ha} |0\rangle.
}}
Next we use \trans\ to relate $b^{\rm out}$ to $b^{\rm in}$.
Consider the left-most oscillator $b_{{\rm R}, \ha}^{\rm out}$.
The only oscillator it can possibly contract with is
$b_{ {\rm R}, \ha}^{{\rm in} \dagger}$ all the way on the right.
This gives a factor of $R(\ha \beta)$.  Following this logic gives
a unique set of non-zero contractions which leads to
\eqn\formtwo{
\langle k_1 | S | k_2 \rangle = \delta_{k_1, k_2}
\prod_{j=1}^{k_1} R((j - \half)\beta) R^*((-j + \half)\beta).
}

Next consider $k_1 , -k_2 \ge 0$.  In this case we 
simply exchange the L and R subscripts on all of the $b^{\rm in}$'s
in \formone.   The contractions then give the transmission coefficient
$T$ instead of $R$.  Therefore \formtwo\ is extended to all
integers $k_1,k_2$ by
writing
\eqn\aaa{
\langle k_1 | S | k_2 \rangle = \delta_{k_1, k_2}
\prod_{j=1}^{|k|} R((j - \half)\beta) R^*((-j + \half)\beta)
+ \delta_{k_1, -k_2} \prod_{j=1}^{|k|}
T((j - \half)\beta) T^*((-j + \half)\beta).
}
This is the complete answer in the discretized theory.

In the continuum limit all of the arguments of $R$ and $T$ go to
zero, and
we are left with
\eqn\vacs{
\langle k_1 | S | k_2 \rangle = {\delta_{k_1,k_2}
\over (1 + e^{-2 \pi \alpha \mu})^{|k|}}
+ { \delta_{k_1, -k_2} \over (1 + e^{+2 \pi \alpha \mu})^{|k|}},
}
where we used \randtsq.

We went through this calculation in exquisite detail simply to
show how the procedure works, but of course the answer \vacs\ could
have been written down immediately after looking at Fig.~5.
First, it is clear that $|k\rangle$ can only mix with $|{\pm}k\rangle$
(in the discretized theory)
by energy conservation.    The first term in \vacs\ corresponds to
$|2 k|$  particles and holes reflecting off of the barrier, while
in the second term all of the particles and holes tunnel through
the barrier, causing $|k\rangle$ to evolve into $|{-}k\rangle$.

Our first indication that the 0B theory suffers classic
infrared problems is that
the $S$-matrix \vacs\ is not unitary when restricted to the
$k$-vacua. 
This is because there are infinitely many additional states
whose energies vanish in the continuum limit and which mix
with the $k$-vacua.  In particular,
the initial state $|1\rangle$ can evolve to the following four
final states:
\eqn\fhlksfjad{\eqalign{
\langle 1 | S | 1\rangle &= R(\half \beta) R^*(-\half \beta) =
{1 \over 1 + e^{-2 \pi \alpha \mu}},\cr
\langle -1|S|1\rangle &= T(\half \beta) T^*(-\half \beta) =
 {1 \over 1 + e^{+ 2 \pi \alpha \mu}},\cr
\langle 0 | a_{{\rm L}, 1} S|1\rangle
&=R(\half \beta) T^*(-\half \beta) =
i { e^{- \pi \alpha \mu} \over 1 + e^{-2 \pi \alpha \mu}},\cr
\langle 0 | a_{{\rm R}, 1} S|1\rangle&=
T(\half \beta) R^*(-\half \beta)=
-i { e^{-\pi \alpha \mu} \over 1 + e^{-2 \pi \alpha \mu}},
}}
where in the final expressions we have taken $\beta \to 0$ inside the arguments.
These are of course nothing more than our old friends from
Fig.~4!  Of course if we
square the amplitudes in \fhlksfjad\ and sum them up we find that
the unitarity condition for $|1\rangle$ is satisfied:
\eqn\aaa{
1 = \sum_{ | \alpha \rangle} | \langle \alpha | S |  1 \rangle|^2.
}

In the continuum limit, we leave the value $k=1$ for the initial state
fixed, and as a consequence the bosons $a_{{\rm L}, 1}$, 
$a_{{\rm R},1}$ in the last
two processes become soft quanta with vanishing energy.  Just like in
QED, a finite fraction of the probability is lost to states with
additional excitations of arbitrarily low energy.  The $S$-matrix is
therefore not well-defined in the continuum limit and we must content
ourselves with transition probabilities.

Let us therefore calculate the total probability for
the initial state $|k\rangle$ to evolve into any state of
charge $l$.
This kind of calculation would be very difficult
in the bosonic formulation because we would have to sum over final
states containing arbitrarily large numbers of $T_{\rm L}$ and $T_{\rm R}$
quanta.  Fortunately the calculation is absolutely trivial in the
free fermion representation.
First note that since the energy of $|k\rangle$ is
$\ha \beta k^2$, energy conservation requires $|l| \le |k|$.
Next note that
the fermionic $S$-matrix \SB\ never mixes $b$ with $b^\dagger$ and
never touches the mode index---all it can do is switch L and
R with probability $\approx |T(0)|^2$ (again setting $\beta = 0$
inside the argument).
Therefore, if we start with
\eqn\aaa{
|{\rm in} \rangle =
|k\rangle = b_{{\rm L}, -k + \ha} \cdots b_{ {\rm L}, - \ha}
b_{{\rm R}, k - \ha}^\dagger \cdots b_{{\rm R}, \ha}^\dagger |0\rangle,
}
then there are only $4^{|k|}$ possible out states given by
\eqn\aaa{
|{\rm out}\rangle = |k\rangle = b_{{\rm F}_1, -k + \ha}
\cdots b_{ {\rm F}_k, - \ha}
b_{{\rm G}_1, k - \ha}^\dagger \cdots b_{{\rm G}_k, \ha}^\dagger |0\rangle,
}
where $\{ F_i, G_i \} \in \{ {\rm L}, {\rm R} \}$.
Furthermore, if we recall the definition of the charge
\eqn\aaa{
C_0 |{\rm out}\rangle = \sqrt{2} m =
{1 \over \sqrt{2}} ( \# b_{\rm L} - \# b_{\rm L}^\dagger
- \# b_{\rm R} + \# b_{\rm R}^\dagger),
}
then there are precisely ${2 |k| \choose |k + m|}$ different
out states for any fixed value of $m \in [ -k, k]$.
Finally note that the probability $|\langle {\rm out}| {\rm in}\rangle |^2$
is actually independent of the particular values of $F_i$, $G_i$ in
the continuum limit (because all energies $\omega$ are taken to the
Fermi sea).  Therefore, we conclude that
the transition probability from the $|k\rangle$-vacuum to
the charge $m$ sector is
\eqn\probtwo{
P( |k \rangle \to {\cal{H}}_m) = \sum_{|\alpha \rangle \in
{\cal{H}}_m} | \langle \alpha| S | k \rangle|^2 =
{ 2 |k| \choose |k + m|}
{ e^{-2 \pi \alpha \mu |k - m|} \over
(1 + e^{-2 \pi \alpha \mu})^{2 |k|}}.
}
Again, this result immediately follows from Fig.~5.
The combinatorial factor ${2 |k| \choose |k + m|}$ counts
how many distinct ways there are for $|k-m|$ particles and holes
to tunnel through the barrier.  This factor multiplies
the transition probability $|R(0)|^{2 |k+m|} |T(0)|^{2|k - m|}$.
We conclude this analysis by noting that the unitarity condition
\eqn\aaa{
1 = \sum_{m=-\infty}^\infty  P (|k\rangle \to {\cal{H}}_m) \qquad
\forall \; k \in \IZ
}
is of course satisfied by \probtwo.

The transition probabilities we calculated in \probtwo\ are notably
undemocratic in the continuum limit.  In particular, $|k\rangle$
can only decay to those $|m\rangle$ with $|m| \le |k|$.  Of course
the origin of this fact is that in the discretized theory, the energy
of $|k\rangle$ is equal to $\beta k^2$ so no process can increase $|k|$.
At first though one might have expected the probabilities
\probtwo\ to depend on $\beta$ in such a way that in the $\beta \to 0$
limit, $P(|k\rangle \to {\cal{H}}_m)$ would depend only on $|k - m|$ and
democracy between the $k$-vacua is restored. However, the order of limits that we take prevents this from happening.
It would be interesting
to determine if there is an alternate regularization procedure
which does give democratic amplitudes in the continuum limit.

\subsec{${\cal{H}}_0 \to {\cal{H}}_0$ amplitudes}

Next we consider amplitudes for processes with no
incoming or outgoing solitons.
This section is brief because in this subspace, the $S$-matrix
is well defined, even in the continuum theory.
For example, the simplest $1 \to 1$ $S$-matrix elements are
\eqn\aaa{
\eqalign{
\langle 0|a_{\rm L}(\omega')  S a_{\rm L}(-\omega)
|0\rangle &=
\langle 0|a_{\rm R}(\omega')  S a_{\rm R}(-\omega)
|0\rangle
= \delta(\omega - \omega')
 \int_0^\omega d\xi\ R(\omega - \xi) R^*(-\xi),\cr
\langle 0| a_{\rm L}(\omega') S a_{\rm R}(-\omega)
|0\rangle &=
\langle 0| a_{\rm R}(\omega') S a_{\rm L}(-\omega)|0
\rangle = \delta(\omega - \omega')
 \int_0^\omega d\xi\ T(\omega - \xi) T^*(-\xi).
}}
Discretization is necessary for the $|k \ne 0\rangle$ vacua because
finite probability is carried away by states with zero
energy, but this does not happen for ${\cal{H}}_0 \to {\cal{H}}_0$
amplitudes for the same reason it did not happen in the 0A theory.

In \DiFrancescoUD, where the Liouville theory on the string
worldsheet was studied, it was shown that correlation functions
of the vertex operators ${\cal{V}}_{\rm L,R}$ (corresponding
respectively to the spacetime fields $T_{\rm L,R}$) factorize at tree
level:
\eqn\facts{
\eqalign{
&\left< \prod_{i=1}^n {\cal{V}}_{\rm L}(k_i)
\prod_{j=1}^m {\cal{V}}_{\rm R}(k_j)\right>= 0, \qquad (n,m \ge 1),\cr
&\left< \prod_{i=1}^n {\cal{V}}_{\rm L}(k_i) \right> = \left< \prod_{i=1}^n
{\cal{V}}_{\rm R}(k_i) \right> = {1 \over 4} \left< \prod_{i=1}^n {\cal{V}}(
\sqrt{2} k_i)\right>_{\rm bosonic},
}}
where ${\cal{V}}$ is the tachyon vertex operator in the bosonic
Liouville theory.
The structure of the exact $S$-matrix in the zero soliton sector
implies that a weaker  form of \facts\ continues to hold at arbitrary
order in string perturbation theory.
Namely, the Liouville theory vertex operator correlation functions continue
to satisfy \facts\ at arbitrary genus after one integrates over the
moduli space.

\subsec{Some ${\cal{H}}_0 \to {\cal{H}}_1$ amplitudes}

Let us consider the amplitude for a simple one-particle
state $|{\rm in} \rangle
\equiv a_{\rm L}(\omega) |0\rangle$ to scatter into the $1$
soliton sector.
In the discretized theory, we write
$a_{\rm L}(\omega)|0\rangle = a_{{\rm L}, n}|0\rangle$, with
the understanding that we will take $n \to \infty$ as $\beta \to 0$
with $\omega = n \beta$ fixed.

The subtleties inherent to massless fields are immediately apparent
when we try to identify possible final states for this process.
The most obvious candidate is $|{\rm out}\rangle
= a_{\rm L}(\omega) |1\rangle=
a_{{\rm L}, n} |1\rangle$, but this cannot mix with $|{\rm in}
\rangle$ in the discretized theory because of energy conservation!
Instead we are forced to consider $a_{{\rm L}, n- 1} |1\rangle$,
which is essentially indistinguishable from $a_{\rm L}(\omega)|1
\rangle$ in the continuum limit.

The $S$-matrix element is
\eqn\aaa{
\langle  1 | a_{ {\rm L}, n' - 1}
S
a_{ {\rm L}, -n}
|0\rangle = T(\half \beta) R^*((\half - n)\beta) \delta_{n,n'}.
}
In the continuum limit this gives
\eqn\aamp{
\langle 1 | a_{\rm L}(\omega') S a_{\rm L}(-\omega)
|0\rangle
= \beta \delta(\omega - \omega') T(0) R^*(-\omega),
}
where the factor of $\beta$ comes from
writing $\delta_{n,n'} = \beta \delta(\omega - \omega')$.
The overall factor of $\beta$ naively implies that
this amplitude can be dropped relative to 
${\cal{H}}_0 \to {\cal{H}}_0$ $S$-matrix
\eqn\relative{
\langle 0 | a_{\rm L}(\omega') S a_{\rm L}(-\omega)
|0\rangle
=  \delta(\omega - \omega') \int_0^\omega d\xi\ R(\omega-\xi)
R^*(-\xi)
}
in the continuum limit.
This conclusion is wrong, however, for reasons that are familiar
by now:
there are infinitely many final states in the charge 1 sector
which give amplitudes that seemingly vanish in the $\beta \to 0$ limit.
Examples include
\eqn\aaa{
a_{{\rm  L}, -n + 1} a_{{\rm L}, -1}|1\rangle, \qquad
a_{{\rm L}, -n + 7} a_{{\rm R}, -3} a_{{\rm L}, -4} |1\rangle, \qquad
{\rm etc.},
}
where, in the continuum limit, $n \to \infty$ but the indices
on the additional quanta are held fixed, rendering them soft.
If we were powerful enough to calculate all of these amplitudes
and sum of their squares, we would find a finite probability
for $a_{\rm L}(\omega)$ to decay into the charge 1 sector.
Fortunately, the power of the free fermion representation
actually makes this calculation trivial! 
The result is
\eqn\sdlfj{\eqalign{
P( a_{\rm L}(-\omega)|0\rangle \to {\cal{H}}_0) &=
{1 \over \omega} \int_0^\omega d\xi\ |R(\omega-\xi) R^*(-\xi)|^2 +
|T(\omega-\xi) T^*(-\xi)|^2,\cr
P(a_{\rm L}(-\omega) |0\rangle \to {\cal{H}}_1) &=
{1 \over \omega} \int_0^\omega d\xi\ |T(\omega-\xi) R^*(-\xi)|^2,\cr
P(a_{\rm L}(-\omega) |0\rangle \to {\cal{H}}_{-1}) &=
{1 \over \omega} \int_0^\infty d\xi\ |R(\omega-\xi) T^*(-\xi)|^2.
}}
Naturally these probabilities add up to one.
The first line is a sum of finite contributions, one of
which comes from the square of \relative.  However the second
and third lines come from summing over an infinite number of
amplitudes such as \aamp\ which individually are all zero
but in total contribute a finite probability.
We note as a curiosity that it is actually possible
to perform the integrals \sdlfj\ analytically.

\subsec{Some ${\cal{H}}_1 \to {\cal{H}}_1$ amplitudes}

Next we consider
\eqn\leading{
\langle 1 | a_{\rm L}(\omega') S a_{\rm L}(-\omega)|1\rangle
= \delta(\omega-\omega') |R(0)|^2 \int_0^\omega d\xi\ R(\omega-\xi)
R^*(-\xi)
+ {\cal{O}}(\beta).
}
The factor of $|R(0)|^2$ comes from contracting the incoming
and outgoing $b_{{\rm R}, \ha}^\dagger$ which appears in the definition
of $|1\rangle$.  Terms in which the $b_{{\rm L}, -\ha}$ in $|1\rangle$
contract with one of the fermions in $a_{{\rm L}}(\omega)$ are
subleading in the continuum limit.

In this case it is consistent to drop the additional ${\cal{O}}(\beta)$
piece.  It cannot possibly contribute a finite amount to transition
probabilities, since if it did, then the leading term in \leading\ would
contribute  an infinite amount of probability, thereby violating unitarity!

\subsec{General structure of the $S$-matrix}

The examples we investigated in the previous subsections illustrate
the general structure of the $S$-matrix.
Decomposing $S$ into the various soliton sectors, we find
\eqn\fullS{
S \sim \pmatrix{& & \vdots & & \cr & |R(0)|^2 S_0 + {\cal{O}}(\beta)
 & {\cal{O}}(\beta) & |T(0)|^2 S_0 + {\cal{O}}(\beta) &  \cr
\cdots & {\cal{O}}(\beta) & S_0 & {\cal{O}}(\beta) & \cdots \cr
&|T(0)|^2 S_0+
{\cal{O}}(\beta) & {\cal{O}}(\beta) & |R(0)|^2 S_0 + {\cal{O}}(\beta) & \cr
& & \vdots & &  }.
}

Here $S_0$ is the perfectly well-defined (although non-unitary),
$\beta$-independent $S$-matrix for scattering in the charge 0 sector.
In other words, $S_0$ is just the `theory II' $S$-matrix of
\MooreZV.
The formula \fullS\ demonstrates schematically that (1)
the $S$-matrix is rendered unitary by including solitonic sectors
labeled by integers $k$ dual to the zero mode of the RR scalar $C$,
and that (2)  as is familiar in massless theories, unitarity forbids
us from dropping amplitudes which naively vanish as the IR cutoff
$\beta$ is taken to zero, since these amplitudes can make
finite contributions to probabilities.

\subsec{Decaying D-branes}

The 0B string theory has unstable D0-branes which have been
interpreted \refs{\TakayanagiSM,\DouglasUP} in the matrix model
as a single fermion perched at the top of the potential (see
Fig.~6).

\fig{Matrix model description of a single unstable 0B D-brane.
The extra hole is present because we have decreed that
we are working in a superselection sector where the total
fermion number is fixed to zero (relative to the total Fermi sea).}
{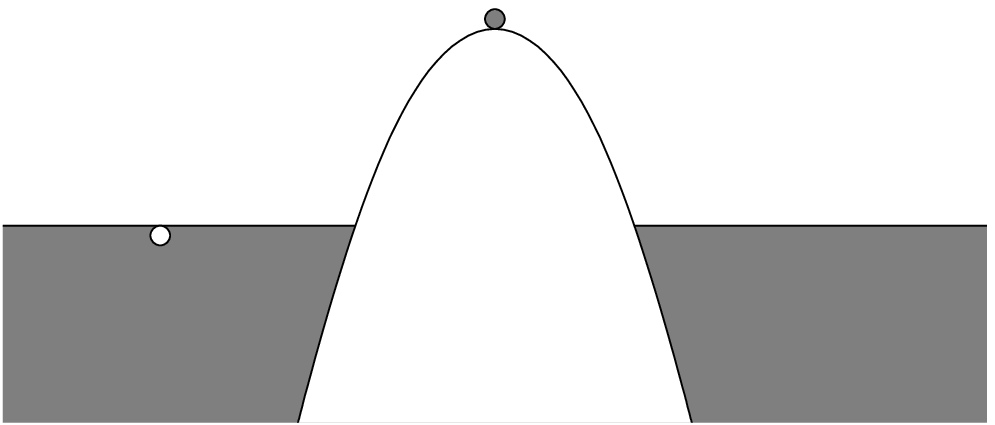}{5.0cm}

Of course these D-branes don't appear as external states in our
$S$-matrix since they are not stable, but we can consider a process
where we send in a very finely tuned initial state specially designed
to kick a single fermion up to the top of the potential.
A state which does the trick is
\eqn\aaa{
|D\rangle = b_{{\rm L},-\hat{\mu}} b_{{\rm L},-\ha}^\dagger |0\rangle.
}
Following our standard conventions, we will keep
$\mu  = \hat{\mu}/\beta$  fixed in the continuum limit.
This state takes a very complicated form when expressed in terms
of bosonic quanta, indicating the large extent to which it is fine-tuned.

Clearly the possible decay products have charge $\pm 1$ or $0$.
A calculation similar to several we have already done
yields the probabilities
\eqn\aaa{\eqalign{
P( |D\rangle \to {\cal{H}}_0) &= {1 \over 2}, \cr
P( |D \rangle \to {\cal{H}}_{-1}) &= {1 \over 2} {1 \over 1+e^{-2 \pi
\alpha \mu}},\cr
P( |D \rangle \to {\cal{H}}_{1}) &= {1 \over 2} { 1 \over 1+e^{+2 \pi
\alpha \mu}}.
}}
Of course these results are completely in line with the physical
picture of Fig.~6.

It is straightforward to calculate $S$-matrix elements for
$|D \rangle$ decay into any particular state in the charge 0 sector.
However $S$-matrix elements into the charge $\pm 1$ sectors will
suffer from the usual IR problems.

\newsec{Conclusion and Discussion}

We have obtained prescriptions for calculating the exact, non-perturbative $S$-matrices for
the ``linear dilaton" vacua of two-dimensional 0A and 0B string theory.
The underlying assumptions are that the proposed string/matrix model dualities
\refs{\TakayanagiSM,\DouglasUP} are correct even non-perturbatively, and that 
the string theory $S$-matrix factorizes into the leg-pole factors
\legB\ times the free-fermion $S$-matrix convolved with bosonization, as in \MPR.
We have shown that in the 0B theory a crucial role is played
by the RR-charged ``soliton" sectors $|k\rangle$, and that infrared
problems require special attention and prevent in some cases the definition of amplitudes,
forcing one instead to work with probabilities.

There are a number of interesting problems to which this technology can be applied.  One 
application is the diagonalization of the Hamiltonian, as was alluded to in the text.  The straightforward expectation is that the eigenstates are the ``theta-vacua" for the states $|k \rangle$,
namely the eigenvectors $| c \rangle$ of the RR scalar zero mode, with a potential $\cos c$.  However this is only obtained
straightforwardly if the $k$-vacua are indistinguishable, something the calculations we have presented argue against.
One may ask whether an alternate regularization procedure than the one
presented here might produce democratic amplitudes.
Further investigation to obtain the correct eigenvectors would naturally be very interesting.

Generalizing these methods to other string theory backgrounds is of course also of interest.  In particular, the two-dimensional type II superstring has been studied in \McGreevyDN; it is claimed to preserve a cubic potential, rather than a quadratic one, in the double scaling limit.
One may consider scattering in this background using our methods; the Fermi theory will no longer be trivial due to the more complicated potential.  The well around the minimum may have an interpretation as a black hole \McGreevyTA.

Another application of $S$-matrix calculations is to determine a low-energy effective action.
One certainly may find solutions of zero-dimensional string theory in the free
fermion picture by just making `blobs' of Fermi sea and letting them evolve through phase space in time, but an effective action provides a bridge to a familiar spacetime interpretation.
However, as discussed in the introduction, at present we do not possess even a lowest-order solution to a lowest-order effective action around which to start expanding.  The $S$-matrices are background-dependent and are only useful once the background around which one is scattering is established.   It would be very useful to determine this lowest-order, linear-dilaton-type solution.

A
non-perturbative description of a string theory also allows one
to address questions about
gravitational physics, such as black holes.  Unstable phenomena such as unstable branes or black holes that form and Hawking decay should appear as resonances in the scattering matrix.  We have discussed briefly how to engineer a state that appears as an unstable D-brane at intermediate times; it would be fascinating to improve our knowledge of brane and gravitational physics by studying further the possible resonances.

Finally, with the exact $S$-matrix in hand it will be possible
to explore quantitatively the strong coupling behavior of
string theory in these two dimensional type 0 toy models (see 
for instance \KapustinHI\ for interesting work in this direction).
A tantalizing speculation concerns the existence of a 
three-dimensional `0M' theory as strong coupling dual of the
0A theory, whose field content can be matched to that of 
3d gravity coupled to a scalar field.

\bigbreak\bigskip\bigskip\centerline{\bf Acknowledgments}\nobreak

We are grateful to D.~Gross, I.~Klebanov, F.~Larsen, J.~McGreevy, S.~Murthy, J.~Polchinski, L.~Rastelli, S.~Shenker and H.~Verlinde for helpful discussions.  
We are particularly thankful to G.~Moore for an inspiring remark on bosonization, and to J.~Maldacena for bringing \CallanUB\ to our attention. This work was supported in part by the National Science Foundation under Grants PHY99-07949 (OD, MS, AV, JW),
0243690 (OD) and PHY00-98395 (RR), as well as by the DOE under Grant
No.~91ER40618 (RR).  Any opinions, findings, and conclusions or
recommendations expressed in this material are those of the authors
and do not necessarily reflect the views of the National Science
Foundation.

\listrefs

\end